\documentclass[preprint,showpacs,preprintnumbers,amsmath,amssymb]{revtex4-2}
\usepackage{bm}
\usepackage{amsmath}
\usepackage{amssymb}
\usepackage{amsfonts}
\usepackage{graphicx}
\begin{document}

\title{ \textbf{
    Spin-State Teleportation and Tests of EPR Correlations Using 151 MeV
    Entangled Protons
} }

\author{H.~Wita{\l}a}
\affiliation{M. Smoluchowski Institute of Physics, 
Faculty of Physics, Astronomy and Applied Computer Science,
Jagiellonian University, PL-30348 Krak\'ow, Poland 
}

\date{\today}

\begin{abstract}
  We discuss the feasibility of quantum spin-state teleportation in a
  three-proton system at an energy of $151$~MeV, where, similarly to the
  low-energy case, a single Bell-state term dominates the proton-proton
  scattering matrix.

  We find that, in contrast to the low-energy regime, where unpolarized
  proton-proton scattering produces strongly entangled outgoing proton pairs,
  at higher energies a pair of protons, each with an energy of $151$~MeV,
  must be produced in an unpolarized, exclusive proton-deuteron breakup
  reaction under complete final-state-interaction kinematics. The subsequent
  interaction of one of the entangled protons with a polarized hydrogen
  target triggers, as in the low-energy case, the teleportation process,
  whereby the polarization of the target proton is transferred to the second
  member of the entangled pair within the very narrow angular region of
  strong entanglement centered around a laboratory scattering
  angle of $45^\circ$.

  We also find that scattering one member of a strongly correlated proton
  pair forming a Bell state from an unpolarized hydrogen target leads
  to Einstein–Podolsky–Rosen-like correlations between the polarization
  of the scattered proton and that of the unscattered second entangled proton.

  The polarization of the scattered proton varies with its scattering angle.
  An identical polarization, following the angular dependence of the
  polarization of the first proton, is induced in the second member of
  the pair, whose initial polarization was zero and whose momentum remains
  unchanged. The sign of the polarization of the second proton is determined
  by the sign of the spin correlation in the Bell state.
\end{abstract}


\maketitle \setcounter{page}{1}

\section{Introduction}
\label{intro}

Recently, experimental verification of the intriguing possibility of
teleporting a quantum-mechanical state between two protons within a
three-proton system has been proposed at low proton
energies \cite{wit_unp_pd,wit_tel_low}. Such an experiment, originally
proposed in \cite{teleport}, is now being seriously considered due to
the availability of strongly entangled $pp$ states formed in unpolarized
proton-proton ($pp$) scattering at energies
below $\approx 20$~MeV \cite{wit_unp_pd,shen_2025}, offering an additional test
of Einstein-Podolsky-Rosen (EPR) correlations \cite{epr}.

The experiment is essentially based on two elements. The first is
the formation of an entangled $pp$ pair, and the second is the scattering
of one of the entangled protons off a polarized hydrogen target. The latter
process is responsible for teleporting the quantum spin state of the target
proton to the second entangled proton, leaving the two remaining protons
in a new entangled state.

For the experiment to be successful, both the production of a specific
Bell-type entangled state of two protons and the actual occurrence of
teleportation require a specific single-term structure of the transition
matrix in both $pp$ scattering processes, with the dominant
contribution originating from a particular Bell component.

In \cite{wit_tel_low}, we investigated quantum-state teleportation at
low laboratory energies of the incoming proton by numerically simulating
the proposed experimental setup using a realistic
nucleon–nucleon (NN) interaction.

We demonstrated the presence of strong teleportation signatures when
a polarized hydrogen target is employed over a broad range of
target polarizations.

The proposed experimental setup requires a polarized hydrogen target,
which currently represents an insurmountable obstacle to its practical implementation at low energies. Excessive energy losses of protons in
currently available polarized targets \cite{watanabe,tateishi} require
their optimization before such experiments can be performed.

This prompted us to consider the feasibility of a teleportation experiment
with higher-energy protons, for which energy losses would be drastically
reduced. The possibility of using higher-energy protons in such
an experiment is provided by the strong dominance of
a single Bell component, $| \psi^+ \rangle \langle \phi^- |$, 
in the $pp$ scattering matrix at an incoming proton energy restricted to
a very narrow band centered around $151$~MeV and in a scattering-angle
region close to  $\theta_{lab}=45^\circ$ \cite{wit_unp_pd,shen_2025}.

The energy of the entangled protons produced in such
a scattering process, $\approx 75.5$~MeV, falls outside the energy region
in which a single Bell component dominates the $pp$ scattering matrix,
thereby precluding their use in the teleportation experiment. The scattering
of an entangled proton from the hydrogen target in the teleportation
experiment requires its energy to be around $151$~MeV.

Since high-quality entangled Bell states   $| \psi^- \rangle$
can be formed in an exclusive proton-deuteron ($pd$) breakup in the $pp$
final-state-interaction (FSI(pp)) kinematically complete geometry, with
an appropriate choice of the incoming proton energy \cite{wit_unp_pd},
we investigate the feasibility of a teleportation experiment using an
entangled pp pair produced in this way.

The entangled pair of protons in the Bell state  $| \psi^+ \rangle$, formed
in $151$~MeV unpolarized $pp$ scattering, provides another fascinating
opportunity to directly test EPR correlations, in a way that is even
simpler than the teleportation experiment, as it does not
require a polarized target.

Namely, scattering one of the entangled protons from an unpolarized hydrogen
target proton would induce a polarization of the outgoing proton.
Due to the strong correlations present in the initial entangled Bell state,
this polarization should be reflected in the polarization of the unscattered
second entangled proton. In particular, the initial polarization of this proton
vanishes because it is a member of a strongly correlated pair forming
the Bell state. After the scattering of the first entangled proton
from the hydrogen target, however, the second proton should acquire
a polarization equal in magnitude to the induced polarization of
the first proton, with its sign determined by the type of Bell state
and its magnitude additionally depending on the scattering angle of
the first entangled proton.

In Sec.\ref{formNN}, for the convenience of the reader, we first briefly
outline the basics of the scattering formalism for beam and target states
prepared in specific spin configurations, together with the main results
of Refs.\cite{wit_unp_pd,wit_tel_low} relevant to the present investigation.
We then in Sec.~\ref{simul} present and discuss
the proposed teleportation experiment with entangled protons, each having
an energy of $151$~MeV. The section concludes with a presentation and
discussion in Sec.~\ref{eprtests} of the experiment designed to
test EPR correlations.

In Appendix~\ref{a1}, the standard spin formalism of nuclear physics
is applied to both experiments, and expressions for the final polarizations
and correlations are derived.

Section~\ref{sumary} contains a summary and conclusions.

\section{Three Proton System with Entangled Initial States}
\label{formNN}

In what follows, we will investigate feasibility of the teleportation
of the quantum mechanical spin state in a three-proton system
occuring at higher energies of participating protons.
This  phenomenon is directly connected
with a novel feature arising when working with entangled $pp$ Bell
states, expressed by  strong correlation between the proton spins and reflected
in a maximal value of the spin-correlation coefficient of $\pm 1$. This is
in contrast with standard scattering experiments, where the polarizations
of the participating particles in the initial state are prepared in a
completely independent manner.

Such large value of the spin correlation imply that the results of
spin-projection measurements performed on the two protons are perfectly
correlated: the result of a spin measurement on one proton uniquely
determines the outcome of the corresponding measurement on the other
proton. This leads to interesting consequences when one of the entangled
protons is scattered off a polarized hydrogen target at energies and
scattering angles for which a single Bell term dominates the $pp$
scattering matrix.

Under such condition, the final spin state of the second proton in the Bell
pair becomes closely related, with identical or opposite polarization value,
to the spin state of the hydrogen target, while
the scattered proton pair emerges in a Bell state of the type, which,
similarily as the sign of the second proton polarization, depends on the 
dominating Bell term in the scattering matrix. 
This process may be interpreted as the teleportation of
the quantum-mechanical spin state from the hydrogen target to the initially
entangled proton.

A representative example of maximal entanglement is given by
the Bell-state basis \cite{bookqinf}, defined as:
\begin{eqnarray}
  | \psi_1\rangle  &\equiv& | \phi^+ \rangle = \frac {1} {\sqrt{2}}
  (| +\frac {1} {2} +\frac {1} {2}\rangle  +
   | -\frac {1} {2} -\frac {1} {2}\rangle ) \equiv
 \frac {1} {\sqrt{2}} (| ++ \rangle  + | -- \rangle )  \cr
  | \psi_{2}\rangle  &\equiv&  | \phi^- \rangle = \frac {1} {\sqrt{2}}
  ( | +\frac {1} {2} +\frac {1} {2}\rangle  -
    |-\frac {1} {2} -\frac {1} {2} \rangle  ) \equiv
 \frac {1} {\sqrt{2}} (| ++ \rangle  - | -- \rangle )  \cr
  | \psi_{3}\rangle  &\equiv&  | \psi^+ \rangle = \frac {1} {\sqrt{2}}
  ( | +\frac {1} {2} -\frac {1} {2}\rangle  +
    |-\frac {1} {2} +\frac {1} {2} \rangle  )  \equiv
 \frac {1} {\sqrt{2}} (| +- \rangle  + | -+ \rangle )  \cr
  | \psi_{4}\rangle  &\equiv&  | \psi^- \rangle = \frac {1} {\sqrt{2}} 
  ( | +\frac {1} {2} -\frac {1} {2}\rangle  -
    |-\frac {1} {2} +\frac {1} {2} \rangle  )  \equiv
 \frac {1} {\sqrt{2}} (| +- \rangle  - | -+ \rangle ) ~.
\label{eq_1}
\end{eqnarray}

The elastic scattering of a proton beam off a proton target is described
by a transition operator $M$ \cite{book}, which can be expressed either
in terms of sixteen complex coefficents 
either in the protons spin-projection basis $\vert m_1 m_2 \rangle$, or
 in the Bell basis (\ref{eq_1}) ~\cite{wit_unp_pd}.

The spin density matrix of the  outgoing protons is given in terms of the
transition operator $M$ and the spin state of the incoming protons, 
described by the density matrix
$\rho_{\text{in}}$, as \cite{ohlsen1972}:
\begin{eqnarray}
  \rho_f &=& M ~ \rho_{in} ~ M^{\dagger}
   ~.
 \label{eq_8}
\end{eqnarray}

In standard $pp$ scattering experiments, the initial spin states of the
proton beam and proton target are prepared independently, leading to
an initial spin density matrix given by the tensor product of the beam
and target spin density matrices. Each of these can be expressed in terms
of the corresponding polarization vectors of the beam, $P_i^{b}$, and
the target, $P_i^{t}$, $ i=1,2,3 ; (x,y,z)$ 
\cite{ohlsen1972}:
\begin{eqnarray}
  \rho_{\text{in}} &=& \frac{1}{4} (I^b + \sum_{i=1}^3 P_i^b \sigma_i)
  \otimes (I^t + \sum_{i=1}^3 P_i^t \sigma_i )    ~,
 \label{eq_8a}
\end{eqnarray}
where  $\sigma_i$ are the standard Pauli matrices.

For an unpolarized proton beam and target, the initial density matrix is
$\rho_{\text{in}} = \frac{1}{4} I^b \otimes I^t$  \cite{ohlsen1972}, and the
final spin density matrix, expressed in the Bell basis (\ref{eq_1}), is
also characterized by sixteen complex coefficients  \cite{wit_unp_pd}.

In Refs.~\cite{wit_unp_pd}, the properties of the $pp$ $M$ matrix
and the final-state spin density matrix $\rho_f$ in unpolarized $pp$
scattering were investigated. It was found that, at the center-of-mass
angle $\theta_{c.m.} = 90^\circ$, only three terms contribute to both
quantities. In addition, a strong dominance of a single term was observed
for both the $M$ matrix and $\rho_f$ at low energies (below around $20$~MeV)
as well as at higher energies (within a narrow region around $151$~MeV).

Specifically, at lower energies, $E_{lab} \approx 10$~MeV, and for angles
around $\theta_{c.m.} \approx 90^\circ$, the transition matrix is well
approximated by
$M \approx | \psi^- \rangle \langle \psi^- |$,
while the corresponding final spin density matrix is given by
$\rho_f \approx | \psi^- \rangle \langle \psi^- |$.

At higher energies, $E_{lab} \approx 151$~MeV, the transition matrix
is instead approximated by 
$M \approx | \psi^+ \rangle \langle \phi^- |$,
and the corresponding final spin density matrix by
$\rho_f \approx | \psi^+ \rangle \langle \psi^+ |$.

The dominance of a single term in $\rho_f$ directly indicates that the
formation of strongly entangled Bell states in unpolarized $pp$ scattering
is possible \cite{wit_unp_pd}.
However, the energies at which such states can be produced are limited
either to the low-energy region (around $10$~MeV) or to a relatively narrow
region around $151$~MeV. Moreover, the angular region in which strongly
entangled Bell states are formed is centered around $\theta_{c.m.} = 90^\circ$.
At low energies, it spans a relatively broad range,
$\theta_{c.m.} \in (55^\circ,125^\circ)$, whereas around $151$~MeV it
becomes strongly restricted to a narrow region of
$\theta_{c.m.} \in (85^\circ,95^\circ)$.
In both energy regions, the highest-purity Bell state is produced at
$\theta_{c.m.} = 90^\circ$. At other angles, the Bell states become
contaminated by contributions from other components, as evidenced
by the nonvanishing polarization of both entangled protons, equal to
a nonzero induced polarization.

The Bell states produced at low energies are of the $|\psi^{-}\rangle$
type, whereas in $pp$ elastic scattering at $151$~MeV the Bell-state
type changes to $|\psi^{+}\rangle$. In both cases, the energies of
the entangled protons are equal to one half of the incident proton energy
at $\theta_{\mathrm{lab}} = 45^\circ $, and become unequal as the scattering
angle deviates from this value.

This reduction in the energies of the outgoing entangled protons makes
$151$~MeV $pp$ scattering an unsuitable candidate for teleportation studies.
Although it produces an entangled proton pair, the energies of the
participating protons fall outside the region of energy where a single Bell term
dominates the $M$-matrix. As a result, the subsequent scattering process of
one of the entangled protons on a polarized hydrogen target,  
required for teleportation to take place, fails to transfer the spin
state of hydrogen target to the second member of the entangled $pp$ pair.

For this reason, we
have restricted  investigation of the teleportation process in
\cite{wit_tel_low} to the low-energy region.
To gain a better understanding of the underlying reaction mechanisms
and the role of spin degrees of freedom, we formulated the problem
within the standard spin formalism commonly used in nuclear physics 
and performed numerical simulations of the teleportation process using a
realistic proton–proton interaction by calculating the corresponding
spin observables.

The analysis of the reactions leading to teleportation at low energies
 provided unambiguous evidence that the dominance of
a single Bell component in the transition matrices $M$ of the two
contributing $pp$ scattering processes is responsible for the occurrence
of teleportation \cite{wit_tel_low}.

The transition matrix $M_{23}$ of the first $pp$ scattering
(see Fig.~\ref{fig1}, where we show the kinematics of scattering of the three
protons) generates
strongly entangled Bell-like states over a broad range of scattering
angles $\theta_{c.m.}^{2}$ and $\theta_{c.m.}^{2'}$. The process of teleportation
itself is, however, directly driven by the transition matrix $M_{12}$
associated with the second scattering, in which proton $2$ interacts
with hydrogen target $1$ \cite{wit_tel_low}.

The dominance of a single Bell component
$ | \psi^- \rangle_{12}~_{12}\langle \psi^- |$ 
in this transition matrix leads
directly to the teleportation of the polarization of target $1$ to
proton $3'$ and to the formation of a strongly correlated $1'2'$ pair.
In the standard spin formalism, this corresponds to a polarization-transfer
coefficient $$K_y^{y'}(1 \rightarrow 3')=1,$$ 
and an induced spin correlation
$$\langle \sigma_y^{1'} \sigma_y^{2'} \rangle^{\mathrm{ind}}=-1,$$ 
showing, that polarization of target $1$ is completely transferred to the
proton $3'$ and that the proton pair $1'2'$ is forming the Bell state
$| \psi^- \rangle$ \cite{wit_tel_low}.

Numerical simulations of the teleportation process have shown that, even for
small values of the polarization of proton $1$, this polarization is
faithfully teleported to proton $3'$ throughout the entire region of
scattering angles $\theta_{c.m.}^{2}$ and $\theta_{c.m.}^{2'}$ characterized
by strong entanglement \cite{wit_tel_low}.

Measurement of the final polarization of proton $3'$ would provide clear
evidence that quantum teleportation at low energies has indeed occurred.
However, the feasibility of such an experiment is hindered by the large
energy losses experienced by low-energy protons in currently available
polarized targets. This prompted us to explore the possibility of performing
the teleportation experiment at energies around $151$~MeV, where
the dominance of a single Bell term in the $pp$ scattering matrix also appears
to provide favorable conditions for teleportation.

\subsection{Spin-State Teleportation  Using 151 MeV
    Entangled Protons}
\label{simul}

Increasing the energy from low values to $151$~MeV leads to a change in
the dominant term in the $pp$ transition matrix $M$, from
$ | \psi^- \rangle \langle \psi^- |$ at low energies to
$| \psi^+ \rangle \langle \phi^- |$ at $151$~MeV.

\begin{figure}
  \includegraphics[bb=0 140 580 785, scale=0.75]{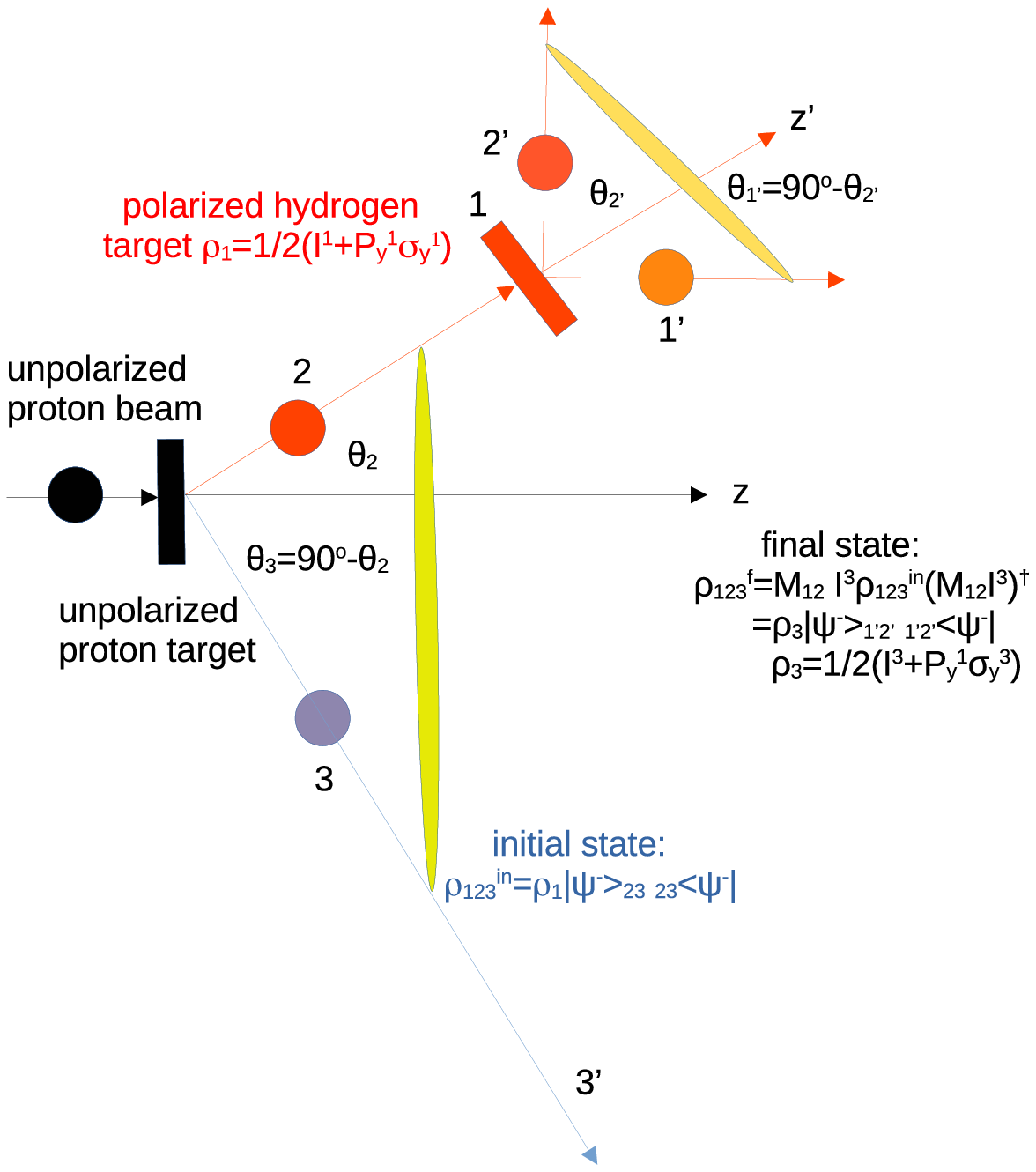}
  \caption{
    Production of low-energy (around $10$ MeV) entangled proton pair $23$ and
    teleportation of the spin state of proton $1$ to proton $3'$ via
    the scattering of proton $2$ off a polarized hydrogen target
    containing proton $1$.
  }
\label{fig1}
\end{figure}

The second difference, related to the fact that the scattered protons
($2$ and $3$, or $1'$ and $2'$) have energies approximately a factor of
two lower than those of the incoming protons at scattering angles
around $\theta_{lab.}=45^\circ$, precludes the feasibility of the
teleportation experiment at $151$~MeV in the arrangement depicted
in Fig.~\ref{fig1}. This is because the dominance of the
$| \psi^+ \rangle \langle \phi^- |$  term in the $pp$ scattering matrix $M$
at $151$~MeV is restricted to a very narrow energy range.

Indeed, when the incoming proton has an energy of $151$~MeV, a strongly
entangled pair $23$ in the Bell state $| \psi^+ \rangle $  is produced.
However, the subsequent scattering of proton $2$ from the hydrogen target
proton $1$ takes place at an energy of approximately $75.5$~MeV, and
the corresponding scattering matrix $M_{12}$ does not satisfy the
teleportation requirement of being dominated by a single Bell term.
Conversely, requiring the energy of proton $2$ to be 151~MeV would require
an incoming proton energy of approximately $302$~MeV, which would
completely destroy the strong entanglement of the pair $23$.

It therefore appears that the only way out of this dilemma is to replace
the first $pp$ scattering in Fig.~\ref{fig1}, whose purpose is to produce
the strongly correlated $pp$ pair $23$ in a Bell state, with a different
process capable of producing such a state.

In Ref.\cite{wit_unp_pd}, it was shown that the unpolarized proton–deuteron
exclusive breakup reaction can produce high-quality Bell states
$ | \psi^- \rangle $ of two protons in kinematically complete
final-state-interaction (FSI) configurations. The availability of such
entangled $pp$ states makes it possible to consider the experimental
arrangement depicted in Fig.\ref{fig2} for exploring the possibility of
spin-state teleportation at $151$~MeV.

In Fig.~\ref{fig2}, the incoming unpolarized proton of sufficiently high
energy interacts with the deuteron target, leading to its breakup,
$d(p,p_2p_3)n$. The two outgoing protons have equal momenta, which is
the exact condition for the $pp$ final-state-interaction (FSI) complete
geometry, with the FSI formation angles satisfying $\theta_2 = \theta_3$. 
This FSI condition uniquely determines the momenta of the three outgoing
nucleons. As shown in \cite{wit_unp_pd}, at each value of the angle $\theta_2$, 
a high-quality Bell state $ | \psi^- \rangle $ of protons $2$ and $3$ is formed.

To demonstrate that, at a sufficiently high incoming proton energy, it is
indeed possible to achieve the pp-FSI complete geometry with the energy
of the protons in the $23$ pair around $151$~MeV, Figs.~\ref{fig3}b) and c)
show the energies and angles of the outgoing nucleons in the
$d(p,p_1p_2)n$  breakup as functions of the FSI production angle
$\theta_1^{lab}=\theta_2^{lab}$ for an incoming proton (or neutron, in the
case of $nd$ breakup) laboratory energy of $E=350$~MeV.
At $\theta_1^{lab}=\theta_2^{lab}=9.1^\circ$,
a Bell state  $ | \psi^- \rangle $  of the proton pair $12$ is produced,
with the two protons having energies  $E_1^{lab}=E_2^{lab}=150.9$~MeV.
The outgoing neutron is scattered at an angle of $\theta_3^{lab}=144.9^\circ$,
with an energy of  $E_3^{lab}=46$~MeV.

To help assess the feasibility of using the proposed experimental arrangement
for proton spin-state teleportation by means of a Monte Carlo simulation,
Fig.~\ref{fig3}a) shows the FSI $pd$ breakup cross section
$d^5\sigma/dS\Omega_1d\Omega_2$ as a function of the FSI production angle
and compares it with the corresponding cross section for the $nd$ breakup
reaction. For details of the three-nucleon Faddeev calculations, with and
without inclusion of the $pp$ Coulomb force, we refer the reader
to \cite{wit_coul}.


Let us now consider the three-proton system consisting of protons
$2$, $3$, and $1$, shown in Fig.\ref{fig2}, with the pair $23$ prepared
as described above, and recall some formulas from Ref.\cite{wit_unp_pd}.

When the $pp$ pair $23$ is in the Bell state $|\psi^-\rangle_{23}$ and
the polarized hydrogen target (proton $1$) has polarization $P_y^1$,
the initial state is described by the spin density matrix~\cite{ohlsen1972}:
\begin{eqnarray}
  \rho_{123}^{\text{in}} =    \rho_{\text{23}} \otimes \rho_{\text{1}} =
 | \psi^- \rangle_{23}   ~_{23}\langle \psi^- |  \otimes 
  \frac{1}{2} ( I^1 + P_y^1 \sigma_y^1 )     
   ~.
 \label{eq_10}
\end{eqnarray}    

The final density matrix of the system after proton $2$ scatters off proton
$1$ is given by:
\begin{eqnarray}
  \rho_f &=& M_{12} \otimes I^3 ~  \rho_{123}^{\text{in}} ~
  (M_{12} \otimes I^3 )^{\dagger}
   ~.
 \label{eq_11}
\end{eqnarray}

Taking the scattering operator in the form
$M_{12}=|\psi^+\rangle_{12}~_{12}\langle\phi^-|$, which is approximately
valid at $E_{lab}\approx151$~MeV, a direct calculation yields 
\cite{wit_unp_pd}:
\begin{eqnarray}
  \rho_f &=&   | \psi^+ \rangle_{1'2'}   ~ _{1'2'}\langle \psi^+ |
  \otimes \frac {1} {2} ( I^3 - P_y^1 \sigma_y^3 )
  ~,
 \label{eq_12}
\end{eqnarray}
demonstrating that the spin state of the target proton $1$ is indeed
teleported to proton $3$, with the polarization $P_y^1$ reversed in sign,
while the scattered proton pair $1'2'$ emerges in a Bell state of the type
$|\psi^+\rangle$.

However, since the above formula was derived under the assumption that only
a single Bell term contributes to the transition matrix $M_{12}$, this
conclusion is strictly valid only, or is best approximated, in the case
where $\theta_{2'}=\theta_{1'}=45^\circ$. In this configuration, the induced
polarization in $pp$ scattering vanishes due to the identity of the protons,
while contributions from additional Bell terms in $M_{12}$ that would spoil
the entanglement are minimized~\cite{wit_unp_pd}. Consequently, the proton
pair $1'2'$ is well approximated by the Bell state $|\psi^+\rangle_{1'2'}$.

For scattering angles $\theta_{2'}$ different from $45^\circ$, nonvanishing
contributions from induced polarizations and polarization-transfer effects
arise due to the increasing importance of additional terms in the $M_{12}$
scattering matrix~\cite{wit_unp_pd}. This leads to nonzero polarizations
of the outgoing protons $1'$ and $2'$. Consequently, the resulting
proton-pair state $1'2'$ can be well approximated by the Bell state
$|\psi^+\rangle_{1'2'}$ only within the very narrow angular range
$\theta_{2'}^{c.m.}\in(85^\circ,95^\circ)$~\cite{wit_unp_pd}.

To investigate how deviations from the dominance of a single Bell term
in the transition matrix $M_{12}$, as well as contamination of
the Bell state by additional contributions, affect the teleportation process,
we performed numerical simulations of the three-proton system by calculating
the final spin density matrix and determining all relevant final-state
spin observables.

To this end, we solved the Lippmann–Schwinger equation for $pp$
scattering~\cite{book} using the high-precision AV18
$NN$ potential~\cite{av18}, with the Coulomb interaction between the two
protons included explicitly. This allowed us to determine the transition
matrices $M_{12}(\theta_{2'})$ required to describe the scattering of
proton $2$ from the target proton $1$ at an incident proton energy of $151$~MeV.

Fig.~\ref{fig4} presents the predictions for the final polarizations
$\langle \sigma_y^{i'} \rangle$ (panels a), c), and e)) and spin correlations
$\langle \sigma_y^{i'}\sigma_y^{j'} \rangle$ ($i',j'=1,2,3$) (panels b), d),
and f)) as functions of the center-of-mass angle $\theta_{c.m.}^{2'}$.
In panels a) and b), the polarization of the hydrogen target is assumed to
be zero, while in panels c) and d) it is set to $P_y^1=0.1$ and
in panels e) and f) to $P_y^1=0.5$.

To better understand the behavior of the observables presented in
Fig.\ref{fig4} and the mechanism of teleportation at $151$~MeV, we
analyzed the three-proton system shown in Fig.\ref{fig2}, following
the standard spin formalism used in nuclear physics. Similar to our
analysis at low energies in Ref.\cite{wit_tel_low}, we express
the final spin observables in terms of induced polarizations, induced
spin correlations, and polarization and spin-correlation transfer
coefficients. In Appendix~\ref{a1}, expressions for the final
polarizations and spin correlations in terms of these observables
are given~\cite{ohlsen1972,wit_spin_doubl,wit_spin_np_entangl}.
We assume that the only polarized particle in the initial configuration
is the hydrogen target proton $1$, with a nonzero polarization component
$P_y^1$, while the proton pair $23$ forms the entangled Bell
state $|\psi^-\rangle_{23}$, with each proton having an energy
of $151$~MeV. For completeness, Appendix\ref{a1} also considers the
case in which the proton pair $23$ is prepared in the Bell
state $|\psi^+\rangle_{23}$.

Assuming further the dominance of a single Bell term,
$|\psi^+\rangle_{12}~_{12}\langle\phi^-|$, in the transition
matrix $M_{12}$, we evaluate the induced polarizations, induced
spin correlations, polarization and spin-correlation transfer
coefficients, as well as the final polarizations and spin correlations
for the three participating protons and all three possible proton pairs.
This provides the corresponding final observables under the
single-Bell-term dominance assumption. Such dominance of a single
term in the transition matrix $M_{12}$ occurs to a very good approximation
at an incident laboratory energy close to $E_{\rm lab}=151$~MeV within
the angular region in which the entanglement is preserved,
$\theta_{2'}^{\rm c.m.}\in(85^\circ,95^\circ)$~\cite{wit_unp_pd}.

Let us first discuss the final spin correlations between different proton
pairs, whose values close to $+1$ provide strong evidence for entanglement
of two protons in the Bell state $|\psi^+\rangle$. It was shown in
Ref.~\cite{wit_unp_pd} that such strongly entangled proton pairs are
indeed formed at energies close to $151$~MeV, and that the dominance of
a single Bell term in the transition matrix $M$ and, consequently, in
the spin density matrix $\rho$, occurs within the very narrow range of
center-of-mass angles $\theta_{2'}^{\rm c.m.}\in(85^\circ,95^\circ)$
(see Fig.28f and Table~I in Ref.\cite{wit_unp_pd}).

The spin correlations are determined by the induced spin correlations
$\langle\sigma_y^{i'}\sigma_y^{j'}\rangle^{\mathrm{ind}}$ (Eq.(\ref{eq_ape9}))
and the single-spin correlation-transfer coefficients
$K_{0y}^{y'y'}(1\to i'j')$ (Eq.(\ref{eq_ape10})) appearing in
Eq.~(\ref{eq_ape8}). For an unpolarized hydrogen target, $P_y^1=0$,
the spin correlations are determined entirely by the induced contribution.

For the proton pair $i'j'=1'2'$, the induced contribution coincides with
that obtained in the scattering of an unpolarized proton $2$ from an
unpolarized hydrogen target proton $1$ (see Eq.(\ref{eq_ape12})).
For this pair, the coefficient $K_{0y}^{y'y'}(1\to i'j')$ is also identical
to its counterpart in unpolarized $pp$ scattering (see Eq.(\ref{eq_ape12.1})).

When a single Bell-state term,
$|\psi^+\rangle_{12}~_{12}\langle\phi^-|$,
dominates in $M_{12}$, the induced spin correlation
$\langle \sigma_y^{1'} \sigma_y^{2'} \rangle^{\mathrm{ind}}$ becomes
$\langle \sigma_y^{1'} \sigma_y^{2'} \rangle^{\mathrm{ind}}=+1$, and the
proton pair $1'2'$ is in the Bell state $|\psi^+\rangle_{1'2'}$. This
behavior occurs in the angular region of strong entanglement, as shown
in Fig.~\ref{fig4}b), where the induced spin correlation
$\langle \sigma_y^{1'} \sigma_y^{2'} \rangle^{\mathrm{ind}}$ approaches
$+1$. This demonstrates that the dominance of a single Bell-state
contribution in the transition matrix $M_{12}$ is responsible for the
formation of the strongly entangled pair $1'2'$ in the state
$|\psi^+\rangle_{1'2'}$.

It is interesting to compare this result with the low-energy case, where
the dominant Bell-state term
$|\psi^-\rangle_{12}~_{12}\langle\psi^-|$
in the scattering matrix $M_{12}$ led to the Bell state
$|\psi^-\rangle_{1'2'}$, characterized by
$\langle \sigma_y^{1'}\sigma_y^{2'}\rangle^{\mathrm{ind}}=-1$
~\cite{wit_unp_pd}.

As can be seen in Fig.~\ref{fig4}b), in this angular region of
$\theta_{2'}^{c.m.}$, the spin correlations of the pairs $2'3'$ and
$1'3'$ take on relatively small and nearly equal values of
$\approx -0.2$. The initially strongly correlated protons $2$ and $3$,
prepared in the Bell state $|\psi^-\rangle_{23}$ with a spin-correlation
coefficient of $-1$, become less correlated after the scattering of
proton $2$ from the hydrogen target $1$, while the initially uncorrelated
protons $3$ and $1$ become slightly correlated.

In the limiting case where only a single Bell-state term contributes to
the scattering matrix $M_{12}$ in this angular region, both of these
spin-correlation coefficients should vanish. The small but nonzero value
of $\approx -0.2$ can therefore be attributed to the contribution of
additional nonvanishing Bell-state components to the matrix $M_{12}$.

A comparison of the spin correlations shown in Figs.\ref{fig4}b), d), and f)
reveals that the polarization of target $1$ has a negligible effect on
the spin correlations in the angular region of strong entanglement. This
can be traced back to the vanishing of the corresponding single-spin
correlation transfer coefficient at $\theta_{2'}^{c.m.}=90^\circ$ and to the
relatively small values of these coefficients at other angles within
the strong-entanglement region (see Fig.\ref{fig5}b)).

Outside the angular region of strong entanglement, a small influence
of $P_y^1$ on the spin correlations of the pairs $1'3'$ and $2'3'$ can
be observed in Figs.\ref{fig4}d) and f) around $\theta_{2'}^{c.m.}=35^\circ$
and $145^\circ$, where the single-spin correlation transfer coefficients
reach their maximum values (see Fig.\ref{fig5}b)).

Turning now to the final polarizations, let us first discuss, in light of
the results obtained in Appendix~\ref{a1}, the final polarizations of
protons $1'$ and $2'$. The induced contribution, $P_y^{1'(2')\mathrm{ind}}$,
to the final polarization of protons $1'$ and $2'$ in Eq.(\ref{eq_ape4}) is,
at each angle $\theta_{c.m.}^{2'}$, given by the induced polarization in
the scattering of proton $2$ from proton $1$
(see Eqs.(\ref{eq_ape7.4}), (\ref{eq_ape7.1}),
and Refs.\cite{ohlsen1972,wit_spin_np_entangl}).

Since the proton pair $23$ is in the state $|\psi^-\rangle_{23}$,
the polarizations of protons $2$ and $3$ vanish. Consequently, the additional
contribution to the final polarizations is determined by
the polarization-transfer coefficients $K_y^{y'}(1 \to 1'(2'))$
(see Eq.(\ref{eq_ape3})). For both protons $1'$ and $2'$, these coefficients
are given by the corresponding polarization-transfer coefficient for
$pp$ scattering at the center-of-mass scattering angle
$\theta_{c.m.}^{2'}$ (see Eqs.(\ref{eq_ape7.5}) and (\ref{eq_ape7.1})).
In both cases, these observables correspond to $pp$ scattering at
the energy of the incoming proton $2$.

For an unpolarized hydrogen target, $P_y^1=0$, the final polarizations
$\langle \sigma_y^{1'} \rangle$ and $\langle \sigma_y^{2'} \rangle$
coincide with the induced polarization in the scattering of proton $2$
from proton $1$. This accounts for the behavior observed in
Fig.~\ref{fig4}a), including its dependence on
$\theta_{c.m.}^{2'}$, which determines the magnitude of the induced
polarization.

For a polarized hydrogen target, the final polarizations
$\langle \sigma_y^{1'} \rangle$ and $\langle \sigma_y^{2'} \rangle$
receive an additional contribution from the polarization-transfer
coefficients $K_y^{y'}(1 \to 1'(2'))$:
$$ Tr (\rho_f ) \langle \sigma_y^{1'(2')} \rangle =  Tr ( \rho_f^0 ) [  
   P_y^{1'(2')~ind} +  P_y^1 K_y^{y'}(1 \to 1'(2')) ]  ~,  $$
leading to modifications of the induced polarizations of protons $1'$
and $2'$ according to the values of the polarization-transfer
coefficients $K_y^{y'}(1 \to 1'(2'))$ shown in Fig.\ref{fig5}a).
These modifications result in the final polarizations shown in
Figs.\ref{fig4}c) and e).

For proton $3'$, the induced contribution to its final polarization
in Eq.(\ref{eq_ape4}), $P_y^{3'\mathrm{ind}}$, given by Eqs.~(\ref{eq_ape7.2})
and (\ref{eq_ape7.1}), is completely determined by the scattering of
proton $2$ from proton $1$.

A direct evaluation of
$Tr ( M_{12} |\psi^-\rangle_{23}~_{23}\langle\psi^-|
M_{12}^{\dagger}\sigma_y^{2'})$ in Eq.(\ref{eq_ape7.4}), which determines
the induced polarization of proton $2'$, using time-reversal invariance
of the NN interaction \cite{book}, yields
$Tr ( M_{12} |\psi^-\rangle_{23}~_{23}\langle\psi^-|
M_{12}^{\dagger}\sigma_y^{3'})$ in Eq.(\ref{eq_ape7.2}), which determines
the induced polarization of proton $3'$, with the opposite sign.

It follows that, at any scattering angle $\theta_{2'}^{c.m.}$, both
inside and outside the angular region of strong entanglement, the induced
polarizations of protons $2'$ and $3'$ are equal in magnitude but
opposite in sign:
$P_y^{3'\mathrm{ind}}=-P_y^{2'\mathrm{ind}}$.
We will return to this interesting point later, in
Sec.~\ref{eprtests}, when discussing the possibility of testing EPR
correlations.

In the angular region of strong entanglement,
$\theta_{c.m.}^{2'} \in (85^\circ,95^\circ)$, the induced contribution
to the final polarization of proton $3'$ should vanish, which is indeed
the case, as shown in Fig.~\ref{fig4}a).

Moreover, in this angular region, the polarization-transfer coefficient
$K_y^{y'}(1 \to 3')$ reaches the value
$K_y^{y'}(1 \to 3')=-1$ (see Appendix~\ref{a1} and Fig.\ref{fig5}a)).
Since the trace appearing in Eq.(\ref{eq_ape7.6}) vanishes,
$$Tr (M_{12}  | \psi^- \rangle_{23}  ~_{23}\langle \psi^- |
\sigma_y^1 M_{12}^{\dagger} ) = 0 ~, $$ 
it follows that
$$Tr(\rho_f) = Tr( \rho_f^0).$$

Consequently, the final polarization of proton $3'$ in the region of
strong entanglement is given by
$$
\langle \sigma_y^{3'} \rangle = - P_y^1.
$$

The condition $K_y^{y'}(1 \to 3')=-1$ provides a clear signature of the
teleportation process and can be directly linked to the dominance of a
single Bell-state contribution
$| \psi^+ \rangle_{12}  ~_{12}\langle    \phi^- |$  
in the transition matrix $M_{12}$,
which describes the scattering of proton $2$ from proton $1$ (see
Eq.~(\ref{eq_ape7.3})).

Polarizing the target proton $1$ therefore modifies the final
polarizations of all three protons (see Figs.\ref{fig4}c) and e)).
For protons $1'$ and $2'$, however, the polarizations remain close to
zero in the angular region of strong entanglement. Since the pair
$1'2'$ emerges in a strongly correlated Bell-like state, the
polarizations of protons $1'$ and $2'$ should approach zero in the
corresponding c.m. angular region. This is indeed the case for an
unpolarized hydrogen target $1$, as shown in Fig.\ref{fig4}a).

However, as the polarization $P_y^1$ increases, the polarizations of
protons $1'$ and $2'$ at $\theta_{2'}^{c.m.}=90^\circ$, which are equal
to each other, exhibit a clear deviation from zero, as seen in
Figs.\ref{fig4}c) and e). This behavior can be attributed to the small
but nonvanishing polarization-transfer coefficients
$K_y^{y'}(1 \to 1')$ and $K_y^{y'}(1 \to 2')$ (see Fig.\ref{fig5}a)),
which arise from the admixture of additional Bell-state components in
the scattering matrix $M_{12}$.

In contrast, the polarization of proton $3'$ changes dramatically in
the angular region of strong entanglement, exhibiting a clear signature
of the teleportation of the spin state of proton $1$ to proton $3'$ (see
Figs.~\ref{fig4}c) and e)). For both values of $P_y^1$, at
$\theta_{c.m.}^{2'}=90^\circ$, the polarization of proton $3'$ is
exactly equal in magnitude to the target polarization $P_y^1$, but has
the opposite sign. Thus, the polarization of target proton $1$ is
completely transferred, with opposite sign, to proton $3'$ over the
narrow angular region of strong entanglement centered around
$\theta_{2'}^{c.m.}=90^\circ$.

It is clear that the most conclusive evidence of teleportation would be
provided by a measurement of the polarization of proton $3'$
at $\theta_{2'}^{c.m.}=90^\circ$ using
a polarized target proton $1$. Owing to the higher energies of the protons
involved, the problem of excessive energy losses should be significantly
reduced, making the feasibility of such an experiment more likely.

\subsection{Tests of EPR Correlations Using Entangled Protons}
\label{eprtests}

As shown in Appendix~\ref{a1}, scattering one proton from a strongly
entangled Bell pair off an unpolarized hydrogen target leads, at any
scattering angle, not only to the polarization of the scattered proton
but also to the simultaneous induction of an identical polarization in
the second proton of the entangled pair, with the sign determined by
the sign of their spin correlation.

This behavior results from the strong spin correlation between the protons
in the Bell state and from the fact that the scattering of an unpolarized
proton from an unpolarized hydrogen target constitutes a measurement
process. As a result of this measurement, the proton scattered at a given
angle becomes polarized. Owing to the strong correlation between the spins
of the two protons, the second proton must acquire the corresponding
polarization, as in the EPR paradox \cite{epr}.

This suggests that, before undertaking the rather complex teleportation
measurement described above, one should first consider an experimental
test of these EPR correlations using an experimental setup without
a polarized hydrogen target and with an unpolarized proton beam.

The possibility of such a test is illustrated by the results shown
in Fig.~\ref{fig4}a), where the induced polarization of proton $3'$
is exactly equal in magnitude but opposite in sign to that of proton $2'$.
This behavior of the polarizations of protons $2'$ and $3'$ results from
the strong spin correlation between protons $2$ and $3$ in the initial
Bell state $|\psi^-\rangle_{23}$, characterized by the spin-correlation
coefficient
$\langle \sigma_y^{2}\sigma_y^{3}\rangle^{\mathrm{ind}}=-1$.

If protons $2$ and $3$ were instead prepared in the initial Bell state
$|\psi^+\rangle_{23}$, the polarization of proton $3'$ would be equal
to that of proton $2'$.

This behavior of the final polarizations for $P_y^1=0$ suggests the
following experimental setup for testing EPR correlations with
an unpolarized hydrogen target.

As in Fig.~\ref{fig1}, an unpolarized proton beam with an energy of
$151$~MeV impinges on an unpolarized proton target, producing, at
scattering angles $\theta_{2}^{lab}=\theta_{3}^{lab}=45^\circ$, a pair of
strongly correlated protons $2$ and $3$, each with an energy of $75.5$~MeV,
in the Bell state $|\psi^+\rangle_{23}$. The pair is characterized
by a spin-correlation coefficient
$\langle \sigma_y^{2}\sigma_y^{3}\rangle^{\mathrm{ind}}=+1$ and vanishing
polarizations of both protons.

A subsequent scattering of proton $2$ from the unpolarized hydrogen
target $1$ leads to the final polarizations of the three outgoing
protons shown in Fig.~\ref{fig6}a). Since in this case
$\langle \sigma_y^{2}\sigma_y^{3}\rangle^{\mathrm{ind}}=+1$, the induced
polarization of proton $3'$, emerging at the scattering angle
$\theta_{3}^{lab}=45^\circ$, should be equal to the induced polarization
of proton $2'$ at the scattering angle $\theta_{2'}^{c.m.}$. These
polarizations can be measured via left–right scattering from
a target with a known proton analyzing power.

The energy of proton $2$, $75.5$~MeV, lies well outside the energy region
in which a single Bell-state term dominates the $pp$ scattering matrix $M$.
This is reflected in the spin-correlation coefficients for all proton pairs,
which generally deviate from $+1$, except at very forward and backward angles
for the pairs $2'3'$ and $1'3'$, respectively (see Fig.\ref{fig6}b)).

The scenario described above is valid at any incoming proton energy
for which a single Bell-state term dominates the scattering matrix $M$,
and therefore the scattered $pp$ pair forms a Bell state. As an example,
at low energies, we consider an incoming proton energy of $10$~MeV in
Fig.~\ref{fig6}c). In this case, protons $2$ and $3$, each with an energy
of $5$~MeV, form a high-quality Bell state $|\psi^-\rangle_{23}$,
characterized by the spin-correlation coefficient
    $\langle \sigma_y^{2}\sigma_y^{3}\rangle^{\mathrm{ind}}=-1$.
    Consequently, the induced polarization of proton $3'$ has the opposite
    sign to that of proton $2'$.

    Since lowering the energy not only improves the quality of the produced
    Bell state but also increases the angular region of strong correlations
    \cite{wit_tel_low}, the second scattering of proton $2$ from the hydrogen
    target $1$ is likewise governed by a single Bell-state term,
    $|\psi^-\rangle_{12}~_{12}\langle\psi^-|$. This leads to a strongly
    correlated proton pair $1'2'$ forming the Bell state
    $|\psi^-\rangle_{1'2'}$ over a wide angular region in
    $\theta_{2'}^{c.m.}$, as shown in Fig.~\ref{fig6}d). Again, the other two
    proton pairs, $1'3'$ and $2'3'$, are only weakly correlated, except at
    very forward and backward angles.

    Unfortunately, due to the very small values of the induced polarizations,
    the low-energy region cannot be used for testing EPR correlations.
    Higher energies, where the induced polarizations are significantly
    larger, are preferable for this purpose. The use of high-quality Bell
    states from FSI($pp$) configurations in exclusive $pd$ breakup, as
    discussed above in the context of teleportation at higher energies, could
    substantially extend the energy range available for experimental tests
    of EPR correlations.

\section{Summary and Conclusions}
\label{sumary}

We investigated the possibility of high-energy teleportation of a
quantum-mechanical spin state in a three-proton system, with the aim of
identifying a simple and experimentally feasible signature of this
process, similar to that found for low-energy teleportation
\cite{wit_tel_low}.

The analysis of the reactions leading to teleportation in a three-proton
system, using the standard spin formalism of nuclear physics, provides
unambiguous evidence that the dominance of a single Bell-state
component in the transition matrix $M_{12}$ for $pp$ scattering, in which
one of the strongly entangled protons scatters from a polarized hydrogen
target, is responsible for the occurrence of teleportation.

Since, at higher energies, the dominance of a single Bell-state term in
the $pp$ scattering matrix $M$ occurs only in a very narrow energy range
around $151$~MeV, teleportation of the spin state of a polarized hydrogen
target requires a Bell state of two protons, each having approximately
this energy. Such a high-quality Bell state $|\psi^-\rangle$ can be
produced in a complete FSI($pp$) configuration of an exclusive
unpolarized $pd$ breakup.

We have shown that the use of such a Bell state leads to teleportation of
the polarization $P_y^1$ of the hydrogen target when one of the strongly
entangled protons scatters within a very narrow angular region of strong
entanglement, centered around the scattering angle
$\theta_{2'}^{lab}=45^\circ$. The other, unscattered member of the
strongly entangled pair becomes polarized with polarization $-P_y^1$.
Measurement of this polarization via left–right scattering from a target
with a known proton analyzing power would provide clear evidence that
quantum spin-state teleportation has occurred.

We have shown that scattering one of two strongly correlated protons forming
a Bell state from an unpolarized hydrogen target leads to an outgoing
proton with an induced polarization that depends on the scattering angle.
Owing to the strong spin correlations in the Bell state, this angular
dependence must be reflected in the polarization of the second, unscattered
member of the entangled pair, which emerges at a fixed angle.

Depending on the type of Bell state, $|\psi^+\rangle$ or $|\psi^-\rangle$,
its polarization must be equal in magnitude and either have the same or
the opposite sign, respectively, to the induced polarization of the scattered
member of the pair.

This provides an opportunity to experimentally test EPR correlations at a
wide range of energies using high-quality $|\psi^-\rangle$ Bell states
produced in complete FSI($pp$) configurations of exclusive
unpolarized $pd$ breakup. However, the most promising scenario
appears to be a test in which the initial Bell state $|\psi^+\rangle$ of
two protons is produced in unpolarized $pp$ scattering at $151$~MeV, with
the strongly entangled protons emerging at a laboratory angle
of $\theta^{lab}=45^\circ$.

\appendix

\section{Final Polarizations and Spin Correlations in a Three-Proton System
  Induced by a  $pp$ Bell State  $|\psi^{\pm} \rangle$}
\label{a1}

\begin{figure}
  \includegraphics[bb=0 292 576 787, scale=0.75]{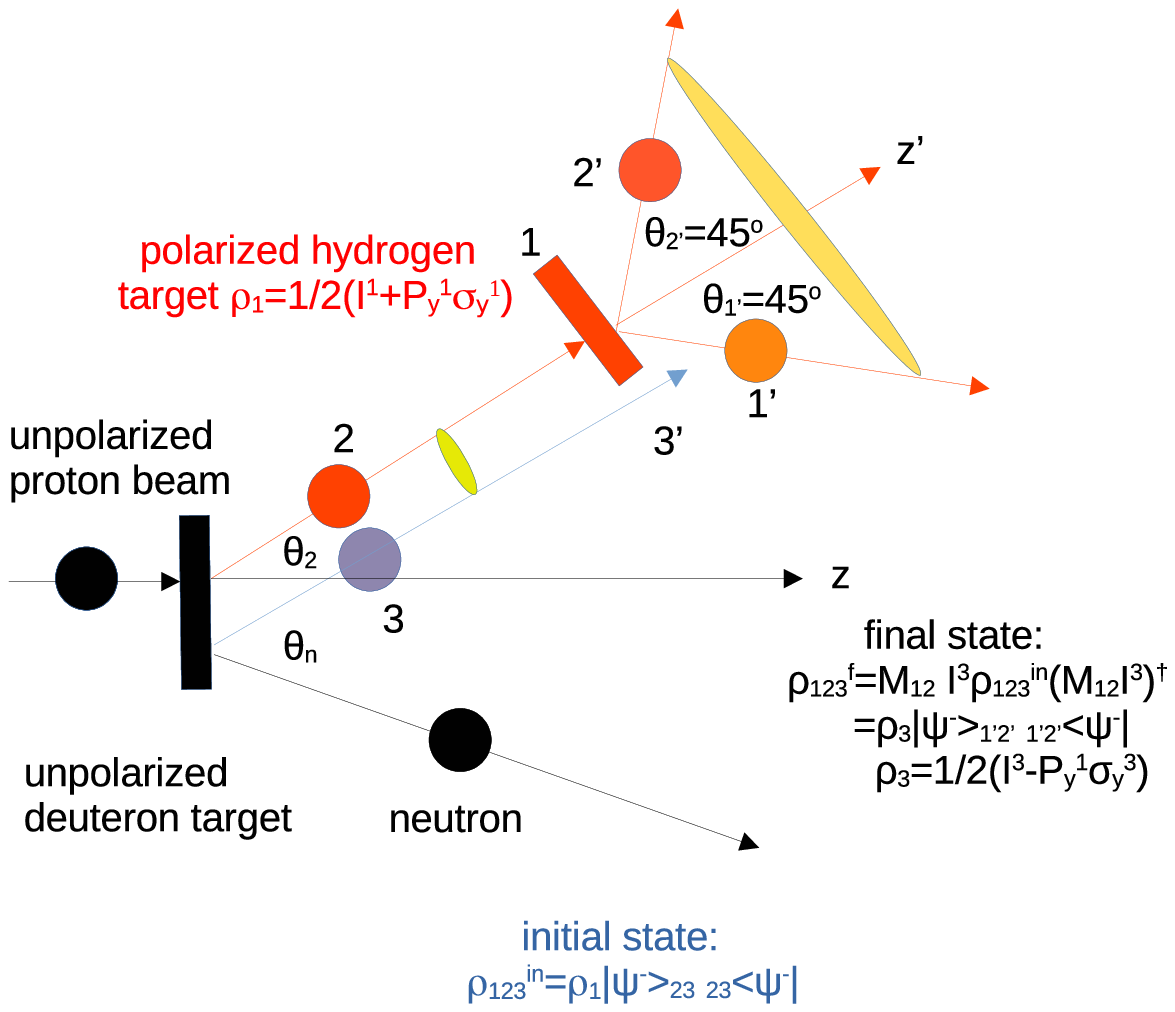}
  \caption{
    Production of a high-energy (about $151$~MeV) strongly entangled proton
    pair $23$ in the Bell state $\lvert \psi^- \rangle_{23}$ through
    unpolarized exclusive proton–deuteron breakup under
    the FSI(pp) condition, followed by teleportation of the spin state of
    proton $1$ to proton $3'$ via scattering of proton $2$ from
    the polarized hydrogen target containing proton $1$.
  }
\label{fig2}
\end{figure}

\begin{figure}
  \includegraphics[scale=0.7]{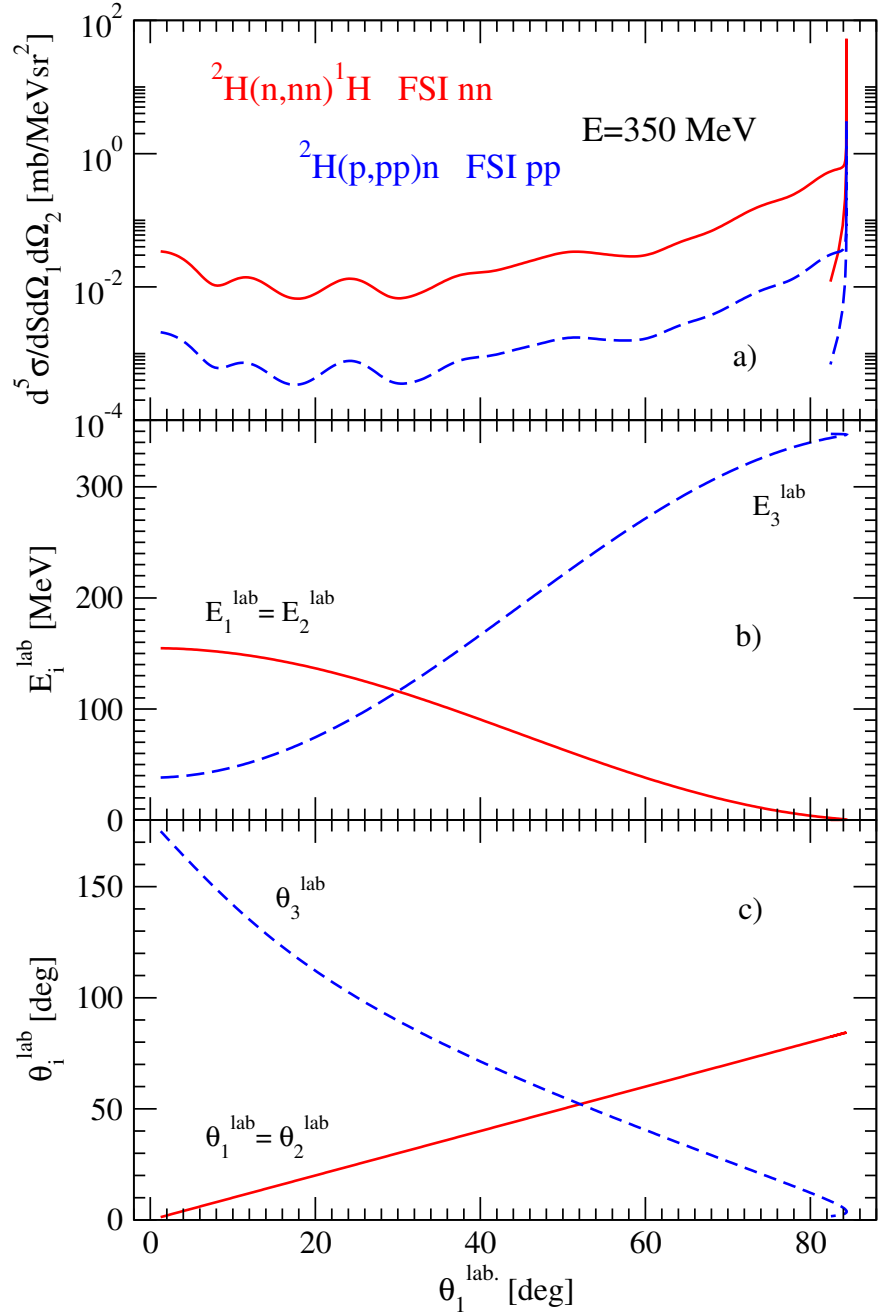}  
\caption{
  (color online)
  The cross section $\frac{d^5\sigma}{dS,d\Omega_1,d\Omega_2}$ for
  the exclusive breakup reactions $^2H(n,nn)^1H$ (red line)
  and $^2H(p,pp)n$ (blue dashed line) at an incident nucleon laboratory
  energy of $E=350$~MeV, under the exact FSI(12) condition, is shown
  as a function of the laboratory angle $\theta_1^{\rm lab}$ in panel a).
  Panels b) and c) show the laboratory energies and angles of
  the outgoing nucleons as functions of $\theta_1^{\rm lab}$,
  using the convention $^2H(N,N_1N_2)N_3$.
  The three-nucleon Faddeev calculations were performed using
  the AV18 $NN$ potential, both with and without the inclusion of
  the Coulomb force, using the $js5j5$ set of partial waves and
  a screening radius of $R_c=20$~fm  \cite{wit_coul}.
  }
\label{fig3}
\end{figure}

\begin{figure}
 \includegraphics[scale=0.74]{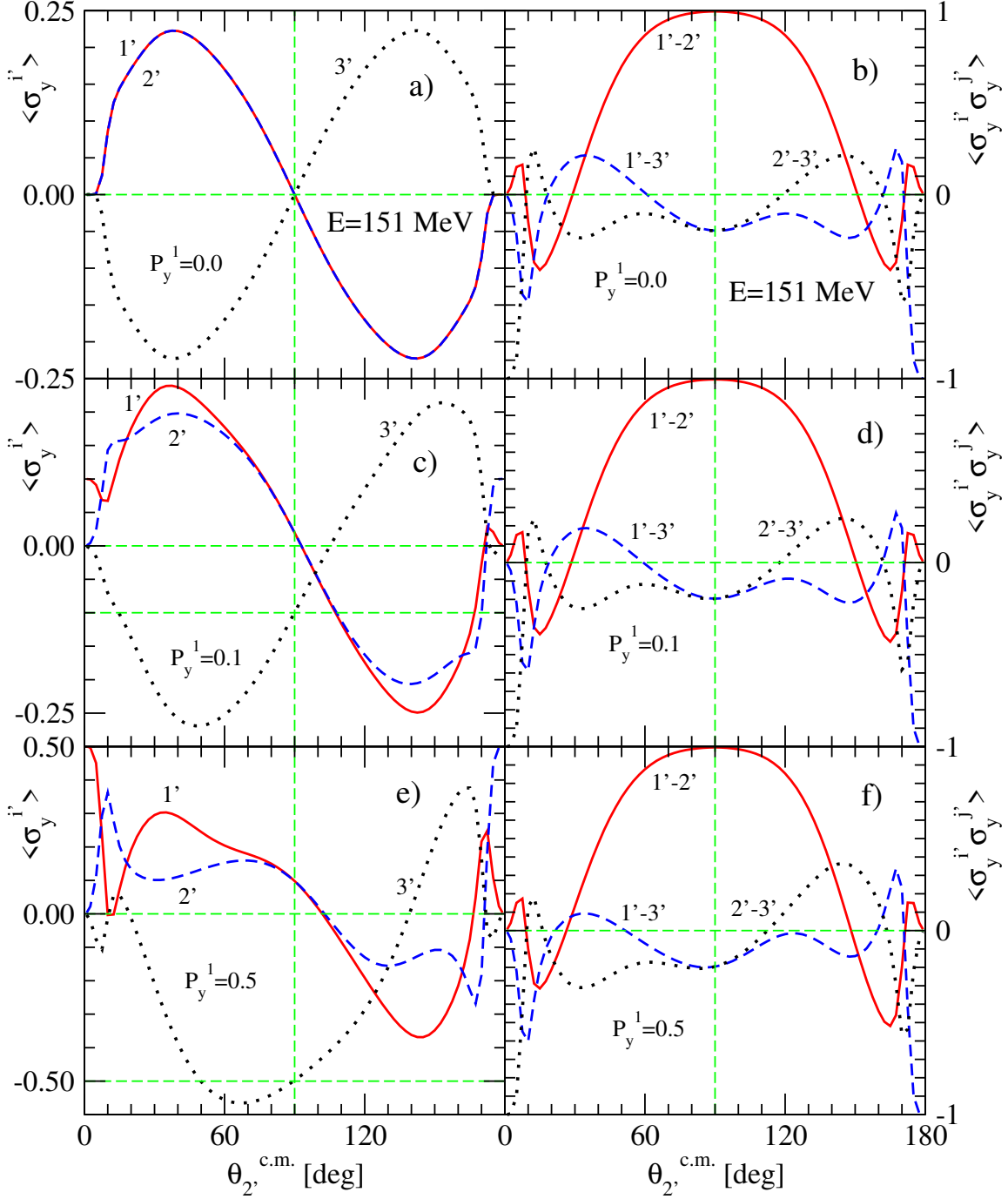} 
\caption{
  (color online)
  Final polarizations $\langle \sigma_y^{i'} \rangle$ of the three
  protons ($i'=1,2,3$) (left panels) and their spin correlations
  $\langle \sigma_y^{i'}\sigma_y^{j'} \rangle$ (right panels) for
  three values of the initial polarization of the hydrogen target
  (proton $1$), $P_y^1$, indicated in each panel. The initial pair of
  entangled protons, each with an energy of $E=151$~MeV and prepared
  in the Bell state $|\psi^-\rangle$, was produced in the $pd$ breakup
  in an exclusive FSI configuration and propagates along the $z'$ axis
  shown in Fig.\ref{fig2}. The calculations were performed using
  the AV18 $NN$ potential and a partial-wave set with $j_{\max}=5$.
   }
\label{fig4}
\end{figure}

\begin{figure}
 \includegraphics[scale=0.8]{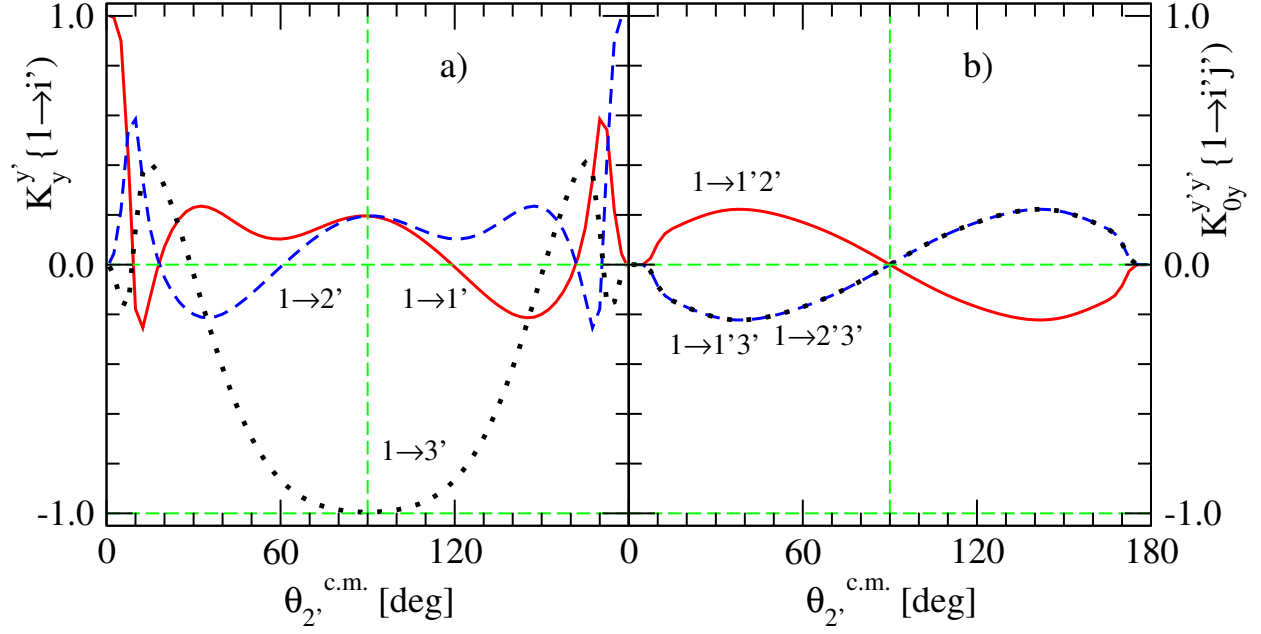} 
\caption{
  (color online)
  Polarization transfer coefficients $K_y^{y'}(1 \to i')$ from proton $1$
  to proton $i'$, $i'=1,2,3$, are shown in panel a), while
  the single-spin correlation transfer coefficients
  $K_{0y}^{y'y'}(1 \to i'j')$ from proton $1$ to proton
  pairs $i'j'$, $i'j'=1'2',1'3',2'3'$, are shown in panel b), as functions
  of the center-of-mass scattering angle $\theta_{2'}^{\rm c.m.}$
  for the three-proton system shown in Fig.~\ref{fig2}.
  The calculations were performed using the AV18 $NN$ potential
  and a partial-wave set with $j_{\max}=5$.
   }
\label{fig5}
\end{figure}

\begin{figure}
 \includegraphics[scale=0.75]{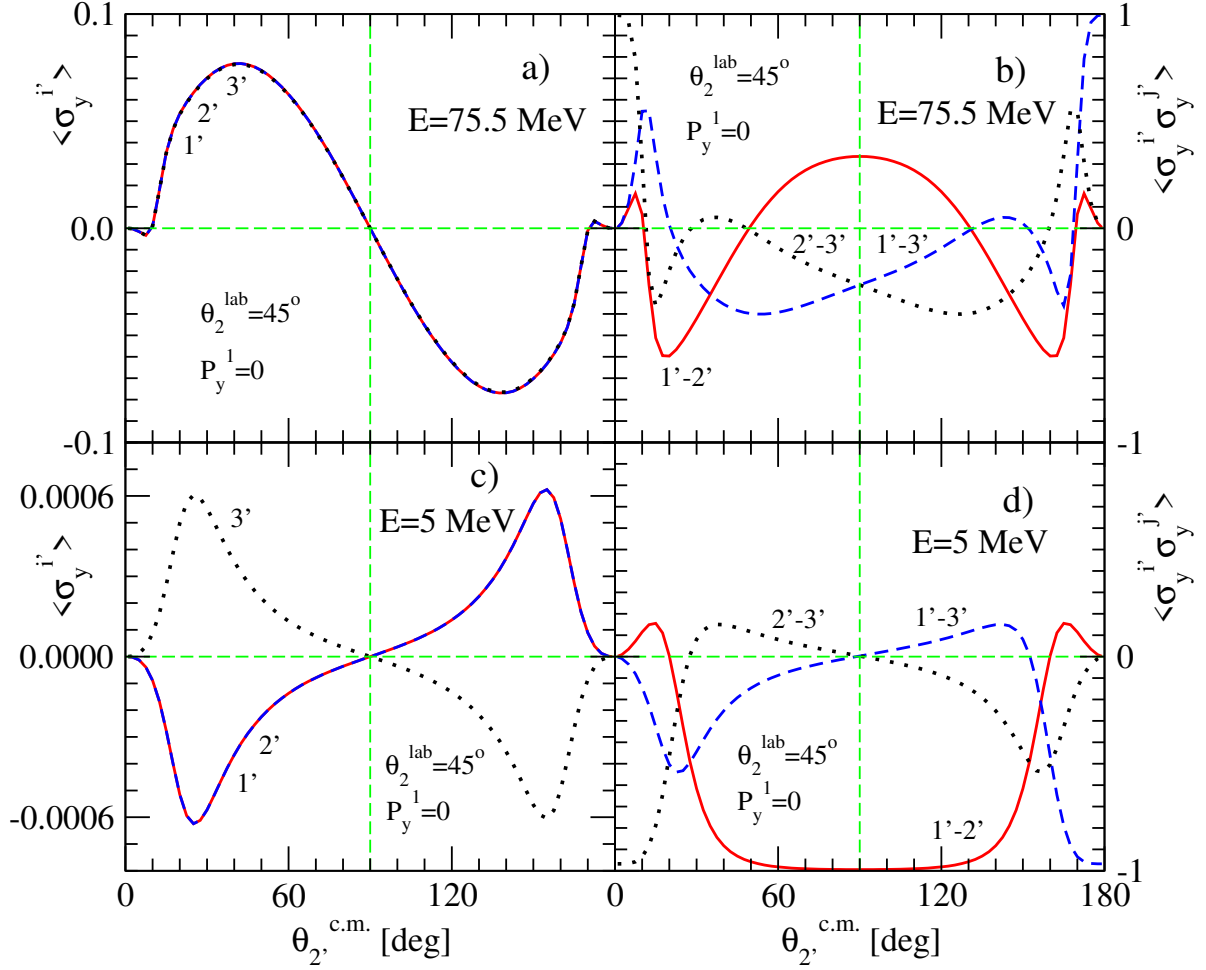} 
\caption{
  (color online)
  Final polarizations $\langle \sigma_y^{i'} \rangle$ of the three
  protons ($i'=1,2,3$) and their spin correlations
  $\langle \sigma_y^{i'}\sigma_y^{j'} \rangle$ for an unpolarized
  hydrogen target (proton $1$), $P_y^1=0$. In panels a) and b),
  the initial pair of entangled protons $2$ and $3$, each with an energy
  of $E=75.5$~MeV and prepared in the Bell state $|\psi^+\rangle$,
  was produced in unpolarized $pp$ scattering with an incident proton
  laboratory energy of $151$~MeV and a laboratory scattering angle
  of $\theta_2^{\rm lab}=45^\circ$ for proton $2$ (see Fig.~\ref{fig1}).
  In panels c) and d),
  the initial pair of entangled protons $2$ and $3$, each with a laboratory
  energy of $E_{\rm lab}=5$~MeV and prepared in the Bell state
  $|\psi^-\rangle$, was produced in analogous unpolarized $pp$ scattering
  with an incident proton energy of $10$~MeV.
  The calculations were performed using the AV18 $NN$ potential and
  a partial-wave set with $j_{\max}=5$.
   }
\label{fig6}
\end{figure}

The strongly correlated $pp$ pair $23$, depicted in  Fig.~\ref{fig2},
is produced in the exclusive 
breakup of an unpolarized proton-deuteron system in a kinematically complete
configuration under the exact FSI(pp) condition at the laboratory
FSI production angle $\theta_2$. Its spin state is described by
the spin-density matrix $\rho_{23}$  \cite{wit_unp_pd}:
\begin{eqnarray}
  \rho_{\text{23}} &=& | \psi^- \rangle_{23}  ~_{23}\langle    \psi^- | 
   ~.
 \label{eq_ape0}
\end{eqnarray}

The incident proton beam energy and the angle $\theta_2$  are chosen such
that the laboratory energies of protons $2$ and $3$ are both $151$~MeV.
This ensures that the subsequent scattering of proton $2$ from the hydrogen
target proton $1$, resulting in the outgoing proton $2'$ emitted at angles
close to $\theta_{2'} = 45^\circ$, is governed by the transition
matrix $M_{12}$, which is dominated by a single Bell-state term,
as required by the teleportation condition \cite{wit_unp_pd}:
\begin{eqnarray}
  M_{12} &=& C | \psi^+ \rangle_{12}  ~_{12}\langle    \phi^- | 
   ~.
 \label{eq_ape0a}
\end{eqnarray}

Let the spin state of the hydrogen target proton $1$ be described by the
spin-density
matrix $\rho_1$ with polarization vector $\vec P_1 = (0, P_y^1, 0)$.
Consequently, the initial spin-density matrix of the three-proton system, 
depicted in Fig.~\ref{fig2}, takes the form:
\begin{eqnarray}
  \rho_{\text{in}} = \rho_{\text{23}} \otimes \rho_{\text{1}} =
  | \psi^- \rangle_{23}  ~_{23}\langle    \psi^- |  \otimes
  \frac{1}{2} (I^1 + P_y^1 \sigma_y^1) 
   ~.
 \label{eq_ape1}
\end{eqnarray}

In the following, we will also consider the case in which the $pp$ pair $23$
forms the strongly entangled Bell state $ | \psi^+ \rangle_{23} $, as,
for example, the pair produced at $151$~MeV in unpolarized $pp$ scattering
at laboratory scattering angles close to $45^\circ$  \cite{wit_unp_pd}.

The final spin-density matrix of the system, after proton $2$ has scattered
from the hydrogen target proton $1$, is given by:
\begin{eqnarray}
  \rho_f &=& M_{12} \otimes I^3 ~  \rho_{\text{in}} ~   (M_{12} \otimes I^3)^{\dagger}
   ~.
 \label{eq_ape2}
\end{eqnarray}

The final polarization of proton $i'(i'=1, 2, 3)$, denoted by 
$\langle \sigma_y^{i'} \rangle$, 
 is therefore given by:
\begin{eqnarray}
  Tr (\rho_f ) \langle \sigma_y^{i'} \rangle &=&  
  Tr ( \rho_f^0 ) [ P_y^{i'~ind} +  P_y^1 K_y^{y'}(1 \to i') ]
   ~,
 \label{eq_ape3}
\end{eqnarray}
where  $P_y^{i'~ind}$ denotes the induced polarization of proton $i'$:
\begin{eqnarray}
  P_y^{i'~ind} &\equiv& \frac {Tr ( M_{12}
    | \psi^\pm \rangle_{23}  ~_{23}\langle    \psi^\pm |
    M_{12}^{\dagger} \sigma_y^{i'} ) }
  {Tr (M_{12}^{\dagger}M_{12}  | \psi^\pm \rangle_{23}  ~_{23}\langle
    \psi^\pm | ) }
   ~,
 \label{eq_ape4}
\end{eqnarray}
and $K_y^{y'}(1 \to i')$ is the polarization-transfer coefficient from
proton $1$ to proton $i'$:
\begin{eqnarray}
  K_y^{y'}(1 \to i') &=& \frac {Tr ( M_{12}
    | \psi^\pm \rangle_{23}  ~_{23}\langle    \psi^\pm |
    \sigma_y^{1} M_{12}^{\dagger} \sigma_y^{i'} ) }
  {Tr (M_{12}^{\dagger}M_{12}  | \psi^\pm \rangle_{23}  ~_{23}\langle
    \psi^\pm | ) }
   ~.
 \label{eq_ape5}
\end{eqnarray}

Here, $\rho_f^0$,  represents the final spin-density matrix corresponding to
an unpolarized target proton $1$ ($P_y^1=0$), and the trace of the final density
matrix is given by:
\begin{eqnarray}
  Tr (\rho_f )  &=&  \frac {1} {2} 
  [ Tr (M_{12} | \psi^\pm \rangle_{23}  ~_{23}\langle    \psi^\pm | M_{12}^{\dagger})
    +  P_y^1 Tr (M_{12} | \psi^\pm \rangle_{23}  ~_{23}\langle    \psi^\pm |
    \sigma_y^1  M_{12}^{\dagger} ) ]
   ~.
 \label{eq_ape3.1}
\end{eqnarray}

Using Eq.~(\ref{eq_1})  for the Bell sta\-tes, one obtains
for the matrix element of $M_{12}$ in (\ref{eq_ape0a}):
\begin{eqnarray}
  \langle m_1 m_2 \vert  M_{12} \vert m'_1 m'_2 \rangle &=& \frac {C} {2}
  ( \delta_{m_1+} \delta_{m'_1+} \delta_{m_2-} \delta_{m'_2+}
  - \delta_{m_1+} \delta_{m'_1-} \delta_{m_2-} \delta_{m'_2-} \cr
&& + \delta_{m_1-} \delta_{m'_1+} \delta_{m_2+} \delta_{m'_2+}
 - \delta_{m_1-} \delta_{m'_1-} \delta_{m_2+} \delta_{m'_2-} )
  ~,
\label{eq_ape6}
\end{eqnarray}
and for the matrix element
$\langle m_2 m_3 \vert   \psi^\pm \rangle_{23}  ~_{23}\langle    \psi^\pm
\vert m'_2 m'_3 \rangle$:
\begin{eqnarray}
  \langle m_2 m_3 \vert  \psi^\pm \rangle_{23}  ~_{23}\langle \psi^\pm
  \vert m'_2 m'_3 \rangle &=& \frac {1} {2}
  ( \delta_{m_2+} \delta_{m'_2+} \delta_{m_3-} \delta_{m'_3-}
  \pm \delta_{m_2+} \delta_{m'_2-} \delta_{m_3-} \delta_{m'_3+} \cr
&& \pm \delta_{m_2-} \delta_{m'_2+} \delta_{m_3+} \delta_{m'_3-}
 + \delta_{m_2-} \delta_{m'_2-} \delta_{m_3+} \delta_{m'_3+} )
  ~.
\label{eq_ape66}
\end{eqnarray}
In the following, we use the shorthand notation $+(-)$ for $+(-)\frac{1}{2}$ 
 to denote the magnetic quantum numbers.

 Substituting the matrix elements of Eq.(\ref{eq_ape66}) into
 Eqs.(\ref{eq_ape4}), (\ref{eq_ape5}), and (\ref{eq_ape3.1}), the
 corresponding traces reduce to:
\begin{eqnarray}
 Tr (M_{12}^{\dagger}M_{12} | \psi^\pm \rangle_{23}  ~_{23}\langle  \psi^\pm |  ) &=&
  \frac {1}  {2} 
          Tr ( M_{12} M_{12}^{\dagger} ) 
 \label{eq_ape7.1}        
\\
Tr ( M_{12}  | \psi^\pm \rangle_{23}  ~_{23}\langle    \psi^\pm |
M_{12}^{\dagger} \sigma_y^{3'})  &=& \pm  \sum_{m_1}
Im ( \langle m_1 - | M_{12}^{\dagger}   M_{12} | m_1 + \rangle )
 \label{eq_ape7.2} 
\\
Tr ( M_{12}  | \psi^\pm \rangle_{23}  ~_{23}\langle    \psi^\pm |
    \sigma_y^1 M_{12}^{\dagger} \sigma_y^{3'} ) &=&
\pm \frac {1}  {2}
[ - \langle + + \vert  M_{12}^{\dagger}M_{12} \vert - - \rangle 
   +\langle + - \vert  M_{12}^{\dagger}M_{12} \vert - + \rangle \cr
 && + \langle - + \vert  M_{12}^{\dagger}M_{12} \vert + - \rangle
   - \langle - - \vert  M_{12}^{\dagger}M_{12} \vert + + \rangle ]
 \label{eq_ape7.3}
   \\
   Tr ( M_{12}  | \psi^\pm \rangle_{23}  ~_{23}\langle  \psi^\pm |
   M_{12}^{\dagger} \sigma_y^{1'(2')})  &=&
\frac {1}  {2} Tr ( M_{12} M_{12}^{\dagger}  \sigma_y^{1'(2')} ) 
 \label{eq_ape7.4}
\\
Tr ( M_{12}  | \psi^\pm \rangle_{23}  ~_{23}\langle    \psi^\pm |
\sigma_y^1 M_{12}^{\dagger} \sigma_y^{1'(2')} ) &=& 
\frac {1}  {2}
Tr ( M_{12} \sigma_y^1 M_{12}^{\dagger} \sigma_y^{1'(2')} ) 
 \label{eq_ape7.5}
\\
Tr (M_{12}  | \psi^\pm \rangle_{23}  ~_{23}\langle \psi^\pm |
\sigma_y^{1} M_{12}^{\dagger} ) &=&
\frac {1}  {2} Tr (  M_{12} \sigma_y^{1} M_{12}^{\dagger} )
\label{eq_ape7.6}
~.
\end{eqnarray}

The last trace is required to evaluate $Tr(\rho_f)$  in Eq.~(\ref{eq_ape3.1}).

Consequently, the induced contribution $P_y^{1'(2')~ind}$ to the final
polarization of protons $1'$ and $2'$ in Eq.~(\ref{eq_ape4}) is equal to
the induced polarization in the  scattering of proton $2$ from proton
$1$ (see Eqs.~(\ref{eq_ape7.4}) and (\ref{eq_ape7.1}), where only 
 the transition matrix $M_{12}$
enters the expression; see also Refs.~\cite{ohlsen1972,wit_spin_np_entangl}).

Similarly, the contribution to the final polarization arising from
the polarization-transfer coefficient  $K_y^{y'}(1 \to 1'(2'))$  is, for
both protons $1'$ and $2'$, equal to the polarization-transfer coefficient
for the scattering of proton $2$ from proton $1$ 
(see Eqs.~(\ref{eq_ape7.5}) and (\ref{eq_ape7.1})).

In both cases, the corresponding observables are those for the scattering of 
proton $2$ from proton $1$,  
evaluated at the energy of the incident proton $2$, which is
entangled with proton $3$.

For proton $3'$, the induced contribution to its final polarization in
Eq.~(\ref{eq_ape3}), $P_y^{3'~\mathrm{ind}}$, given by Eqs.~(\ref{eq_ape7.2})
and (\ref{eq_ape7.1}), is again completely determined by the scattering of 
proton $2$ from proton $1$.
A direct evaluation of
$\frac {1}  {2} Tr ( M_{12} M_{12}^{\dagger}  \sigma_y^{2'} )$ 
in (\ref{eq_ape7.4}), using time reversal invariance of the 
NN interaction \cite{book} (Eq.~(2.43)),
yields (\ref{eq_ape7.2}), with the sign determined by that of the
spin-correlation coefficient
$\langle \sigma_y^{2} \sigma_y^{3} \rangle = \pm 1$ for the Bell
states $|\psi^\pm \rangle$. 
It follows that scattering one proton from a strongly entangled Bell pair
off an unpolarized hydrogen target not only polarizes the scattered proton
at any scattering angle $\theta_{2'}^{c.m.}$, but simultaneously induces an
identical polarization in the second proton, with the sign determined
by the sign of their spin correlation.

When a single Bell term dominates the transition matrix $M_{12}$,
the traces in Eqs.~(\ref{eq_ape7.1})\-(\ref{eq_ape7.6}) can be evaluated
analytically, allowing the corresponding spin observables to be calculated.

 Such dominance occurs, to a very good approximation, at
 $E_{lab} \approx 151$~MeV within the narrow angular range
$\theta_{c.m.}^{2'} \in (85^\circ,95^\circ)$ \cite{wit_unp_pd}. 

 In this region of strong entanglement, the induced contribution to the
 final polarization of proton $3'$ vanishes.

Moreover, in this angular region the polarization-transfer coefficient
$K_y^{y'}(1 \to 3')$ satisfies
$K_y^{y'}(1 \to 3') = - 1$ (see Eqs.~(\ref{eq_ape7.3}) and (\ref{eq_ape7.1}),
together with
$M_{12}^{\dagger}M_{12} = |C|^2 |\phi^-\rangle_{12} ~_{12}\langle\phi^-|$).
Furthermore,
$$
\mathrm{Tr}\left(M_{12}  | \psi^\pm \rangle_{23}  ~_{23}\langle    \psi^\pm |
    \sigma_y^1 M_{12}^{\dagger}\right)=0
$$
as follows from Eq.~(\ref{eq_ape7.6})), implying that 
$\mathrm{Tr}(\rho_f)=\mathrm{Tr}(\rho_f^0)$.
Consequently, in the region of strong entanglement, the final polarization
of proton $3'$  is given by
$$
\langle \sigma_y^{3'} \rangle = - P_y^1.
$$

The result $K_y^{y'}(1 \to 3') = - 1$ demonstrates a teleportation-like
transfer of polarization and can be directly attributed to the dominance of
a single Bell-state component in the transition matrix $M_{12}$, 
which describes the scattering of proton $2$ from proton $1$
(see Eq.~(\ref{eq_ape7.3})). We would like to remind the reader that, at low
energies, where the dominant Bell-state component in the scattering
matrix $M_{12}$ was 
$ | \psi^- \rangle_{12}  ~_{12}\langle    \psi^- | $,  
the polarization-transfer coeﬃcient  $K_y^{y'}(1 \to 3') = + 1$
~\cite{wit_tel_low}.

We now consider the final spin correlations between different protons
following the scattering of proton $2$ from proton $1$. They are given by:
\begin{eqnarray}
  Tr (\rho_f ) \langle \sigma_y^{i'} \sigma_y^{j'} \rangle &=&  
  Tr ( \rho_f^0 ) [ \langle \sigma_y^{i'} \sigma_y^{j'} \rangle^{ind} +
    P_y^1 K_{0y}^{y'y'}(1 \to i'j') ] 
   ~,
 \label{eq_ape8}
\end{eqnarray}
with $i'j'=1'2'$, $1'3'$, $2'3'$.

As in the case of the final polarizations, the spin correlations consist of
two contributions. The first is independent of the polarization of
proton $1$ and is referred to as the induced spin correlation, 
$\langle \sigma_y^{i'} \sigma_y^{j'} \rangle^{ind}$,  for the proton pair
$i'j'$:
\begin{eqnarray}
  \langle \sigma_y^{i'} \sigma_y^{j'} \rangle^{ind} &\equiv& \frac {Tr (M_{12}
    | \psi^\pm \rangle_{23}  ~_{23}\langle    \psi^\pm |
     M_{12}^{\dagger} \sigma_y^{i'} \sigma_y^{j'} ) }
  {Tr (M_{12}^{\dagger}M_{12}   | \psi^\pm \rangle_{23}  ~_{23}\langle  \psi^\pm | ) }
   ~,
 \label{eq_ape9}
\end{eqnarray}
whereas the second is the contribution to the final spin correlation
arising from 
the polarization of proton $1$, namely $P_y^1 K_{0y}^{y'y'}(1 \to i'j')$, where:
\begin{eqnarray}
  K_{0y}^{y'y'}(1 \to i'j') &=& \frac {Tr ( M_{12}
    | \psi^\pm \rangle_{23}  ~_{23}\langle    \psi^\pm |
    \sigma_y^{1} M_{12}^{\dagger} \sigma_y^{i'} \sigma_y^{j'}) }
  {Tr (M_{12}^{\dagger}M_{12}   | \psi^\pm \rangle_{23}  ~_{23}\langle \psi^\pm | ) }
   ~.
 \label{eq_ape10}
\end{eqnarray}
We adopt a convention analogous to that introduced in
Ref.~\cite{wit_spin_np_entangl}, where the corresponding quantity
 $K_{0y}^{y'y'}(1 \to i'j')$ was
referred to as the single-spin correlation transfer coefficient.

Using the matrix elements given in (\ref{eq_ape66}),  
the traces entering Eqs.~(\ref{eq_ape9}) and (\ref{eq_ape10}) 
 for $i'j'=1'2'$ reduce to:
\begin{eqnarray}
  Tr (M_{12}  | \psi^\pm \rangle_{23}  ~_{23}\langle    \psi^\pm |
  M_{12}^{\dagger} \sigma_y^{1'} \sigma_y^{2'} )  &=&
\frac {1} {2} Tr(M_{12} M_{12}^{\dagger} \sigma_y^{1'} \sigma_y^{2'} ) 
~,
\label{eq_ape11}
\\
Tr (M_{12} | \psi^\pm \rangle_{23}  ~_{23}\langle    \psi^\pm |
\sigma_y^{1} M_{12}^{\dagger}
  \sigma_y^{1'} \sigma_y^{2'} )  &=&
  \frac {1} {2} Tr(M_{12} \sigma_y^{1}
  M_{12}^{\dagger} \sigma_y^{1'} \sigma_y^{2'} ) 
   ~.
 \label{eq_ape11.1}
\end{eqnarray}
Consequently, the induced spin correlation
$\langle \sigma_y^{1'} \sigma_y^{2'} \rangle^{\mathrm{ind}}$ is given by
\begin{eqnarray}
  \langle \sigma_y^{1'} \sigma_y^{2'} \rangle^{ind} &=& \frac {Tr (M_{12}
   M_{12}^{\dagger} \sigma_y^{1'} \sigma_y^{2'} ) }
  {Tr (M_{12}^{\dagger} M_{12} ) }
   ~,
 \label{eq_ape12}
\end{eqnarray}
whereas the single-spin correlation transfer coefficient
$K_{0y}^{y'y'}(1 \to 1'2')$ is given by
\begin{eqnarray}
  K_{0y}^{y'y'}(1 \to 1'2') &=& \frac {Tr (M_{12} \sigma_y^{1} 
   M_{12}^{\dagger} \sigma_y^{1'} \sigma_y^{2'} ) }
  {Tr (M_{12}^{\dagger} M_{12} ) }
   ~. 
 \label{eq_ape12.1}
\end{eqnarray}
Both quantities are identical to the corresponding observables for $pp$
scattering, in which proton $2$ scatters from proton $1$.

Assuming, furthermore, that the matrix $M_{12}$ is  dominated by
the single Bell-state component 
$|\psi^+ \rangle \langle \phi^- |$, one obtains
$$\langle \sigma_y^{1'} \sigma_y^{2'} \rangle^{\mathrm{ind}} = +1,$$ 
confirming that the pair $1'2'$ forms the Bell state $|\psi^+ \rangle$. 
In contrast,  $\langle \sigma_y^{1'} \sigma_y^{3'} \rangle^{\mathrm{ind}}$,
$\langle \sigma_y^{2'} \sigma_y^{3'} \rangle^{\mathrm{ind}}$,
as well as all
$K_{0y}^{y'y'}(1 \to i'j')$, vanish.



\acknowledgments

This work was supported by the National Science Centre,
Poland under Grant
IMPRESS-U 2024/06/Y/ST2/00135.   
The numerical calculations were partly performed on the supercomputers of
the JSC, J\"ulich, Germany.




%


\clearpage

\begin{thebibliography}{99}

\bibitem{wit_tel_low} H. Wita{\l}a, archiv:2606.14171v1 [nucl-th].

\bibitem{wit_unp_pd} H. Wita{\l}a, Phys. ReV. C {\bf{113}}, 054001 (2026).

\bibitem{teleport}  B. F. Kostenko et al., archiv:quant-ph/0012133v1.

\bibitem{shen_2025}  Z. X. Shen et al., archiv:2510.24325 [nucl-th].
  
\bibitem{epr} A. Einstein, B. Podolsky, and N. Rosen, 
  Phys. Rev. {\bf 41}, 777 (1935).   

\bibitem{watanabe} A. Watanabe et al., Nucl. Instrum. Methods in Phys. Res.
  A {\bf{ 1078}}, 170562 (2025).
  
\bibitem{tateishi} K. Tateishi et al., arXiv:2508.06549 [physics.ins-det].
  
\bibitem{bookqinf} Michael A. Nielsen and Issac L. Chuang, Quantum
  Computation aand Quantum Information,  Cambridge University Press 2000.

\bibitem{book} W. Gl\"ockle, The Quantum Mechanical Few-Body Problem,
  Springer Verlag 1983.

\bibitem{ohlsen1972} G. G. Ohlsen, Rep. Prog. Phys. {\bf{35}}, 717 (1972).

\bibitem{wit_coul} H. Wita{\l}a, J. Golak, and R. Skibi\'nski, 
  Phys. Rev. C{\bf 110}, 024005 (2024).

\bibitem{av18}  R.B. Wiringa, V.G.J. Stoks, R. Schiavilla,
  Phys. Rev. C{\bf 51}, 38 (1995).

\bibitem{wit_spin_doubl} H. Wita{\l}a, J. Golak, R. Skibi\'nski, H. Sakai,
  K. Sekiguchi, Phys. Rev. C{\bf 111}, 044003 (2025).

\bibitem{wit_spin_np_entangl} H. Wita{\l}a, J. Golak, and R. Skibi\'nski, 
  Phys. Rev. C{\bf 112}, 044002 (2025).
  

  
  
\end{thebibliography}
\end{document}